\documentclass[12pt]{article}
\usepackage{verbatim}
\usepackage{amsfonts}
\usepackage{graphics}
\usepackage{amsmath}
\usepackage{times}
\usepackage{appendix}
\usepackage{color}
\usepackage{soul}
\usepackage{mathtools}
\usepackage{enumerate}
\usepackage{fancyhdr,latexsym,amsmath,amsfonts,amssymb,amsbsy,amsthm,url}
\usepackage[margin=0.5in,footskip=0.25in]{geometry}
\usepackage{graphics,graphicx,epsfig}
\usepackage{bbm}
\usepackage{breqn}
\usepackage{caption,subcaption,float,subfloat,pgfplots}
\usepackage{pdflscape}
\usepackage{hyperref}
\usepackage{tikz}
\usepackage[numbers]{natbib}
\usepackage{setspace}
\usepackage{booktabs}
\usepackage{float}
\usepackage{caption}
\usepackage{titlesec}
\usepackage{times}
\usepackage{multirow}
\usepackage{tikz}
\usepackage{algorithm}
\usepackage{algpseudocode}
\usepackage{subcaption}

\usetikzlibrary{shapes.geometric, arrows.meta, positioning}
\tikzstyle{block} = [rectangle, rounded corners, minimum width=2.8cm, minimum height=1.2cm, text centered, draw=black, fill=gray!10]
\tikzstyle{arrow} = [thick, ->, >=stealth]

\makeatletter

\renewcommand{\section}{
	\@startsection
	{section}
	{1}
	{0pt}
	{1.1\baselineskip}
	{0.2\baselineskip}
	{\sc \centering}
}

\renewcommand{\subsection}{
	\@startsection
	{subsection}
	{1}
	{0pt}
	{1.1\baselineskip}
	{0.2\baselineskip}
	{\sc \centering}
}

\renewcommand{\subsubsection}{
	\@startsection
	{subsubsection}
	{1}
	{0pt}
	{1.1\baselineskip}
	{0.2\baselineskip}
	{\sc \centering}
}

\makeatother

\usepackage[flushleft]{threeparttable}
\usepackage{rotating,booktabs,multirow}
\usepackage{colortbl}
\usepackage{makecell,cellspace,caption}

\begin{document}
	
\title{\large\sc Machine Learning Based Stress Testing Framework for Indian Financial Market Portfolios}
\normalsize

\author{
\sc{Vidya Sagar G} \thanks{Department of Mathematics, Indian Institute of Technology Guwahati, Guwahati 781039, India. Email: g.vidya@iitg.ac.in}
\and \sc{Shifat Ali} \thanks{Department of Mathematics, Indian Institute of Technology Guwahati, Guwahati 781039, India. Email: a.shifat@iitg.ac.in}
\and \sc{Siddhartha P. Chakrabarty} \thanks{Department of Mathematics, Indian Institute of Technology Guwahati, Guwahati 781039, India, Email: pratim@iitg.ac.in}
}

\date{}
\maketitle
\begin{abstract}
	
This paper presents a machine learning driven framework for sectoral stress testing in the Indian financial market, focusing on financial services, information technology, energy, consumer goods, and pharmaceuticals. Initially, we address the limitations observed in conventional stress testing through dimensionality reduction and latent factor modeling via Principal Component Analysis and Autoencoders. Building on this, we extend the methodology using Variational Autoencoders, which introduces a probabilistic structure to the latent space. This enables Monte Carlo-based scenario generation, allowing for more nuanced, distribution-aware simulation of stressed market conditions. The proposed framework captures complex non-linear dependencies and supports risk estimation through Value-at-Risk and Expected Shortfall. Together, these pipelines demonstrate the potential of Machine Learning approaches to improve the flexibility, robustness, and realism of financial stress testing.


\end{abstract}

\section{Introduction}

Stress testing is a fundamental component of financial risk management, designed to evaluate the resilience of financial systems or portfolios under adverse market conditions~\cite{roncalli2019handbook}. By simulating extreme but plausible scenarios, stress tests provide insight into potential vulnerabilities, that may not be evident during normal periods of market operation. Regulatory institutions such as the Basel Committee on Banking Supervision (BCBS)~\cite{bcbs2009stress} and the Reserve Bank of India (RBI)~\cite{rbi2023fsr} have emphasized the critical role of stress testing in ensuring financial stability, particularly in the aftermath of global financial crises of 2008. Emerging markets, particularly India, are increasingly exposed to both domestic vulnerabilities and global economic shifts. The Indian economy includes a broad range of sectors, each with distinct risk sensitivities. However, these sectors are also interdependent, creating complex patterns of financial stress. Events such as the Global Financial Crisis and the COVID-19 pandemic have underscored the need for tools that can offer a more nuanced, sector-aware perspective on financial system vulnerabilities.

Traditional stress testing frameworks have used a range of statistical and econometric approaches to estimate risk under adverse conditions. Factor models~\cite{fama1993common}, while useful for decomposing risk into common components, often assume stable linear relationships that may not hold during crises. Univariate methods such as extreme value theory (EVT)~\cite{mcneil2000tail} focus on tail behavior, but ignore dependencies across assets or sectors. Multivariate approaches, including copulas~\cite{nelsen2006copulas} and correlation-based models, attempt to capture these dependencies, but can be sensitive to model choices and often assume static structures~\cite{li2000default}. While econometric models like ARCH~\cite{engle1982arch} and GARCH~\cite{bollerslev1986garch} are widely used for volatility modeling, yet they typically focus on individual time series and may not scale well to large, interconnected systems. These limitations, across assumptions, dimensionality, and adaptability, suggest that traditional models may not fully capture the complexity of modern financial systems, especially in settings with sectoral heterogeneity and time-varying dependencies. In this context, Machine Learning techniques have the ability to learn patterns directly from data, model nonlinear relationships, and represent high-dimensional structures, thereby offering new possibilities for building more flexible and forward-looking stress testing frameworks \cite{petropoulos2022stress}.

To address these shortcomings, we present a machine learning-based framework for stress testing sectoral portfolios within the Indian financial markets. By leveraging dimensionality reduction techniques such as Principal Component Analysis (PCA)~\cite{WOLD198737}, Autoencoders (AEs)~\cite{bank2021autoencoders}, and Variational AEs (VAEs)~\cite{Kingma_2019}, our approach enables the generation of both deterministic and probabilistic stress scenarios. This allows us to model both linear and complex nonlinear relationships among market variables, providing a more nuanced and data-driven understanding of financial risk. These techniques help capture latent structures in financial time series data and simulate stress scenarios that reflect the intricacies of real-world crises. The resultant framework improves risk sensitivity and interpretability, while offering a flexible tool for forward-looking stress testing. The remainder of this paper details the implementation and empirical evaluation of these techniques, illustrating how each method can be used to identify potential vulnerabilities under different stress scenarios.
 
\section{Machine Learning Stress Testing Framework: Three Pipelines}

The goal of this framework is to evaluate the resilience of sectorally diversified portfolios under a variety of stress conditions. We propose a modular, three-stage stress testing pipeline that accommodates both linear and nonlinear modeling approaches. The process begins with the extraction of latent risk factors from historical return data using one of three dimensionality reduction techniques, namely, PCA AEs and VAEs. These latent representations are interpreted as aggregated risk factors~\cite{packham2023} that summarize common variation across assets and sectors. In PCA, these correspond to linear combinations of the input variables, whereas in AE and VAE, they are learned through nonlinear encoding functions.

In the second stage, stress scenarios are generated by modifying the latent space. For PCA and AE, we apply deterministic perturbations to one or more latent dimensions. For VAE, stress scenarios are constructed by sampling from the learned latent distribution, enabling a probabilistic exploration of the risk landscape. These latent factors can be viewed as underlying drivers of market behavior, capturing co-movement patterns that may correspond to macroeconomic influences, sector-wide trends, or shifts in investor sentiment. By perturbing (in the case of PCA or AE) or sampling from the values of these factors (in the case of VAE), we simulate structural shocks to hidden components of systemic risk, enabling us to generate realistic stress scenarios that propagate through the entire portfolio. This interpretation serves as the foundation for our stress testing methodology, guiding how scenarios are constructed and how their impact is evaluated.

\begin{figure}[h]
\centering
\begin{tikzpicture}[node distance=2cm and 1.8cm, font=\small, >=stealth]

\tikzstyle{block} = [rectangle, draw=black, minimum width=1.8cm, minimum height=1.0cm, align=center]
\tikzstyle{arrow} = [->, thick]

\node (input) [block] {Data Collection\\and Preprocessing};

\node (pca) [block, right=of input, yshift=2.9cm] {PCA Dimensionality\\Reduction};
\node (ae)  [block, right=of input] {AE Dimensionality\\Reduction};
\node (vae) [block, right=of input, yshift=-2.9cm] {VAE Dimensionality\\Reduction};

\node (stress) [block, right=of ae, xshift=1.5cm] {Stress Scenario\\Generation};
\node (impact) [block, right=of stress, xshift=0.2cm] {Portfolio Impact\\Analysis};

\draw [arrow] (input) -- (pca)
  node[midway, above, rotate=29] {\scriptsize Daily returns};
\draw [arrow] (input) -- (ae)
  node[midway, above] {\scriptsize Standardized}
  node[midway, below] {\scriptsize returns};
\draw [arrow] (input) -- (vae)
  node[midway, above, rotate=-29] {\scriptsize Standardized}
  node[midway, below, rotate=-29] {\scriptsize returns};

\draw [arrow] (pca) -- (stress)
  node[midway, above, rotate=-26.5] {\scriptsize Extracted Risk Factors (PCs)}
  node[midway, below, rotate=-26.5] {\scriptsize Perturbation of PCs};
\draw [arrow] (ae) -- (stress)
  node[midway, above] {\scriptsize Extracted Latent Factors}
  node[midway, below] {\scriptsize Perturbation of Latents};
\draw [arrow] (vae) -- (stress)
  node[midway, above, rotate=26.5] {\scriptsize Learnt distribution params}
  node[midway, below, rotate=26.5] {\scriptsize Monte Carlo Simulation};

\draw [arrow] (stress) -- (impact)
  node[midway, above] {\scriptsize Reconstructed}
  node[midway, below] {\scriptsize Returns};

\end{tikzpicture}
\caption{Stress Testing Pipeline using PCA, AE, and VAE.}
\label{fig:framework_flowchart}
\end{figure}
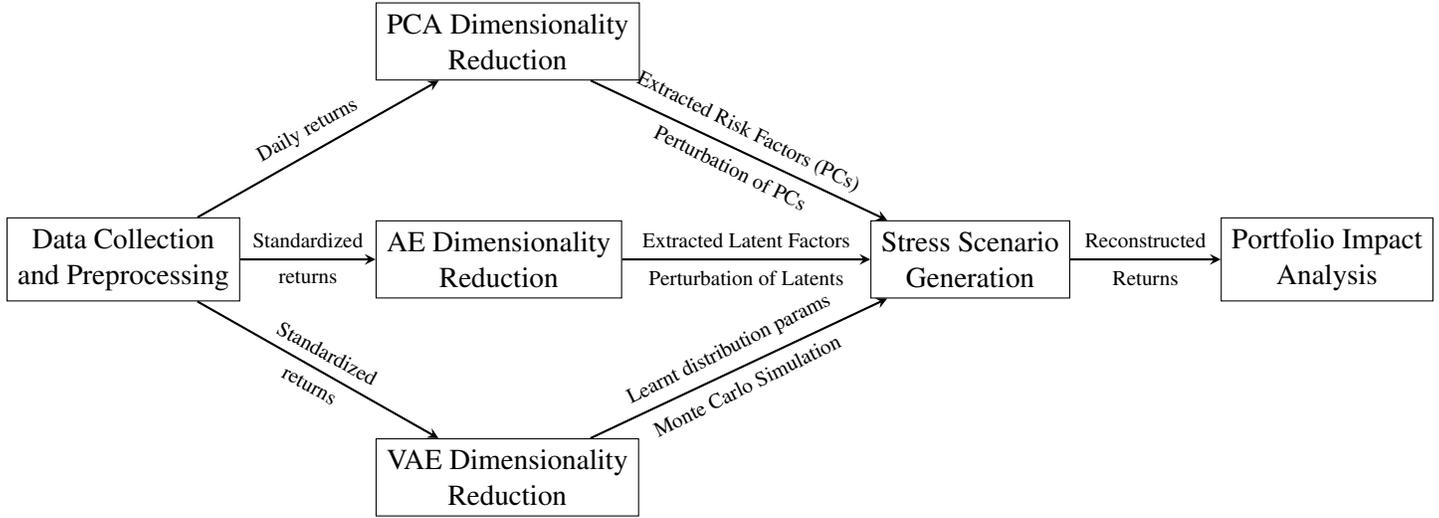

The final stage involves reconstructing the stressed returns from the modified latent representations and evaluating their impact on portfolio-level performance. PCA operates on raw return data, while AE and VAE use standardized returns to enhance model stability and convergence. Portfolio impact is assessed using standard risk metrics including Value-at-Risk (VaR) and Expected Shortfall (ES)~\cite{artzner1999coherent}, and drawdown, with results interpreted both in synthetic stress settings and in the context of real historical market events. This general structure supports both deterministic and probabilistic stress scenario generation and offers flexibility in capturing sector-specific or systemic risk. Figure~\ref{fig:framework_flowchart} illustrates the unified architecture across all three pipelines, which are detailed in the following sub-sections. 

\subsection{PCA-Based Stress Testing}

We implement a rolling-window PCA approach to extract dominant linear risk factors from raw daily return data. For each rolling window, PCA is applied to the asset return matrix to obtain a set of principal components, representing orthogonal directions of maximum variance. These components are interpreted as aggregated risk factors that summarize common patterns of movement across multiple assets or sectors. Instead of relying on predefined economic indicators or hand-crafted stress variables, PCA allows the data to reveal its own structure of risk.

Stress scenarios are created by perturbing the principal components, either individually, to isolate the impact of a single latent factor, or jointly, to simulate multi-factor systemic shocks. Perturbations are applied as fixed multiples of the empirical standard deviation (typically $\pm 2\sigma$), along each component direction. The perturbed latent representation is then projected back into the original asset space using the PCA loading matrix, producing a stressed return vector.

These stressed returns are used to evaluate portfolio-level risk under different scenarios. Portfolios are subjected to the reconstructed stress returns, and metrics such as VaR, ES, and maximum drawdown are computed. The full implementation is summarized in Algorithm~\ref{algo:pca_pipeline}.

\begin{algorithm}[h]
\caption{PCA-Based Stress Testing Pipeline}
\label{algo:pca_pipeline}
\begin{algorithmic}[1]

\For{each rolling window index $t$ of length $w$ over the return matrix $R \in \mathbb{R}^{T \times N}$}
    \State Extract return submatrix $R_t \in \mathbb{R}^{w \times N}$
    \State Fit PCA with $d$ components on $R_t$
    \State Obtain principal components $PC_t \in \mathbb{R}^{d}$ and loading matrix $L \in \mathbb{R}^{N \times d}$

    \For{each stress type in \{Single-Factor, Multi-Factor\}}
        \If{Single-Factor}
            \State Perturb component $PC_i$ by $\pm k \cdot \sigma_i$, keeping other components fixed
        \ElsIf{Multi-Factor}
            \State Apply perturbation vector $\Delta \in \mathbb{R}^{d}$, sampled from $\pm \sigma$
        \EndIf

        \State Reconstruct stressed returns using \texttt{pca.inverse\_transform(PC\_perturbed)}
        
        \State Compute portfolio-level metrics:
        \State \quad Value-at-Risk (VaR)
        \State \quad Expected Shortfall (ES)
        \State \quad Maximum Drawdown

        \State Compute delta metrics:
        \State \quad $\Delta$VaR = VaR$_{\text{stressed}}$ $-$ VaR$_{\text{baseline}}$
        \State \quad $\Delta$ES = ES$_{\text{stressed}}$ $-$ ES$_{\text{baseline}}$
        \State \quad $\Delta$DD = Drawdown$_{\text{stressed}}$ $-$ Drawdown$_{\text{baseline}}$

        \State Compute average sector-level return shifts across assets
    \EndFor
\EndFor

\end{algorithmic}
\end{algorithm}

\subsection{AE-Based Stress Testing}

We implement a rolling-window AE pipeline to extract nonlinear latent risk factors from standardized daily return data. AEs are neural networks trained to reconstruct input data by learning a compressed intermediate representation. These latent vectors are interpreted as nonlinear analogs of aggregated risk factors, capable of capturing complex dependencies and sectoral structure that may be missed by linear methods such as PCA. For the architecture of a typical AE, the interested reader may refer to \cite{wolfram_autoencoder}.

The AE is trained independently on each rolling window of standardized returns using a shallow, fully connected architecture with symmetric encoder and decoder layers. Once trained, the encoder compresses the returns into a $d$-dimensional latent vector. To simulate stress scenarios, individual latent dimensions are perturbed by fixed multiples of their standard deviations (e.g., $\pm 2\sigma$ for single-factor shocks), or jointly perturbed to simulate multi-factor stress. The perturbed latent vector is passed through the decoder to generate stressed returns in the standardized space, which are then inverse-transformed back to the original scale.

The reconstructed stressed returns are evaluated for their portfolio-level impact. Risk metrics such as VaR, ES, and maximum drawdown are computed under each scenario. Delta metrics quantify the deviation from baseline (unstressed) performance, and sector-level effects are assessed by aggregating return shifts within each group. The full pipeline is summarized in Algorithm~\ref{algo:ae_pipeline}.

\begin{algorithm}[h]
\caption{AE-Based Stress Testing Pipeline}
\label{algo:ae_pipeline}
\begin{algorithmic}[1]

\For{each rolling window index $t$ of length $w$ over the standardized return matrix $R \in \mathbb{R}^{T \times N}$}
    \State Train AE on $R_t$ using mean squared error loss and early stopping
    \State Encode input to latent vector $Z_t \in \mathbb{R}^d$
    
    \For{each stress type in \{Single-Factor, Multi-Factor\}}
        \If{Single-Factor}
            \State Perturb latent coordinate $z_i$ by $\pm k\sigma_i$ while holding others fixed
        \ElsIf{Multi-Factor}
            \State Apply vector of perturbations $\Delta \in \mathbb{R}^d$ sampled from $\pm \sigma$
        \EndIf

        \State Decode perturbed latent vector to obtain stressed returns in standardized space
        \State Inverse-transform returns to original scale using stored mean and std

        \State Compute portfolio-level risk metrics:
        \State \quad Value-at-Risk (VaR)
        \State \quad Expected Shortfall (ES)
        \State \quad Maximum Drawdown

        \State Compute delta metrics:
        \State \quad $\Delta$VaR = VaR$_{\text{stressed}}$ $-$ VaR$_{\text{baseline}}$
        \State \quad $\Delta$ES = ES$_{\text{stressed}}$ $-$ ES$_{\text{baseline}}$
        \State \quad $\Delta$DD = Drawdown$_{\text{stressed}}$ $-$ Drawdown$_{\text{baseline}}$

        \State Compute sector-level return shifts by averaging asset-wise changes within each sector
    \EndFor
\EndFor

\end{algorithmic}
\end{algorithm}

\subsection{VAE-Based Stress Testing}

We extend the AE framework by incorporating a probabilistic latent space using a VAE. Unlike standard AEs, VAEs model the latent representation as a distribution rather than a fixed vector, allowing for Monte Carlo-based stress scenario generation. This enables a richer simulation of uncertainty and more diverse stress trajectories through sampling from the learned latent space.
For the architecture of a typical VAE, the interested reader may refer to \cite{vae_blog}.

The VAE is trained on rolling windows of standardized returns using a symmetric encoder-decoder architecture. The encoder learns to parameterize the latent distribution for each sample via a mean vector $\mu \in \mathbb{R}^d$ and a log-variance vector $\log \sigma^2 \in \mathbb{R}^d$. During training, latent vectors are sampled using the reparameterization approach to enable back-propagation through the stochastic layer. To generate stress scenarios, we draw samples from the learned latent distribution and pass them through the decoder to obtain stressed return vectors. Latent variables are sampled using the reparameterization approach, which allows gradients to flow through the sampling step during training. Given the mean $\mu$ and standard deviation $\sigma$ of the learned distribution, the latent vector $z$ is sampled as:
\begin{equation}
z = \mu + \sigma \odot \epsilon, \quad \epsilon \sim \mathcal{N}(0,I),
\label{eq:reparam}
\end{equation}
where $\odot$ denotes element-wise multiplication. This formulation enables stochastic sampling while maintaining differentiability. The full implementation is summarized in Algorithm~\ref{algo:vae_pipeline}.

\begin{algorithm}[h]
\caption{VAE-Based Stress Testing Pipeline}
\label{algo:vae_pipeline}
\begin{algorithmic}[1]

\For{each rolling window $t$ of length $w$ over the standardized return matrix $R \in \mathbb{R}^{T \times N}$}
    \State Train VAE using MSE + KL divergence loss
    \State Encode returns to latent distribution: mean $\mu_t \in \mathbb{R}^d$, log-variance $\log \sigma^2_t \in \mathbb{R}^d$
    
    \For{$m = 1$ to $M$}
        \State Sample $z^{(m)} \sim \mathcal{N}(\mu_t, \sigma^2_t)$ using reparameterization
        \State Decode $z^{(m)}$ to get standardized stressed returns $\hat{R}^{(m)}$
        \State Inverse-transform $\hat{R}^{(m)}$ to original return scale
        \State Compute portfolio return $r^{(m)}$ from $\hat{R}^{(m)}$
    \EndFor

    \State Aggregate all $r^{(1)}, \dots, r^{(M)}$ to form empirical return distribution
    \State Visualize using histogram or kernel density estimate (KDE)

\EndFor

\end{algorithmic}
\end{algorithm}

\section{Dataset and Preliminary Analysis}

This analysis uses historical stock price data from five key sectors of the Indian equity market: Financial Services, Information Technology (IT), Consumer Goods, Energy, and Pharmaceuticals. The dataset comprises daily closing prices collected over a 20-year period (2004–2024), retrieved from Yahoo Finance using the \texttt{yfinance} Python library. This long-term daily dataset enables detailed time series analysis and provides sufficient depth for modeling both short-term and structural market behaviors.

For each sector, five companies were selected based on their market capitalization and prominence within the sector, ensuring that the sample reflects the major players influencing sector-level trends. The final list includes actively traded stocks that are representative of sectoral performance. (see Table~\ref{tab:sector_companies}) Daily returns are computed using the percentage change in closing prices. Minor preprocessing steps, such as handling missing values, are also applied to ensure data consistency and quality.

\subsection{Exploratory Data Analysis}

We begin our exploratory analysis by summarizing the statistical properties of the return series. Basic descriptive statistics, including mean return, standard deviation, and skewness, were computed across selected companies to understand their distributional characteristics.
\begin{figure}[h]
    \centering
    \includegraphics[width=0.6\textwidth]{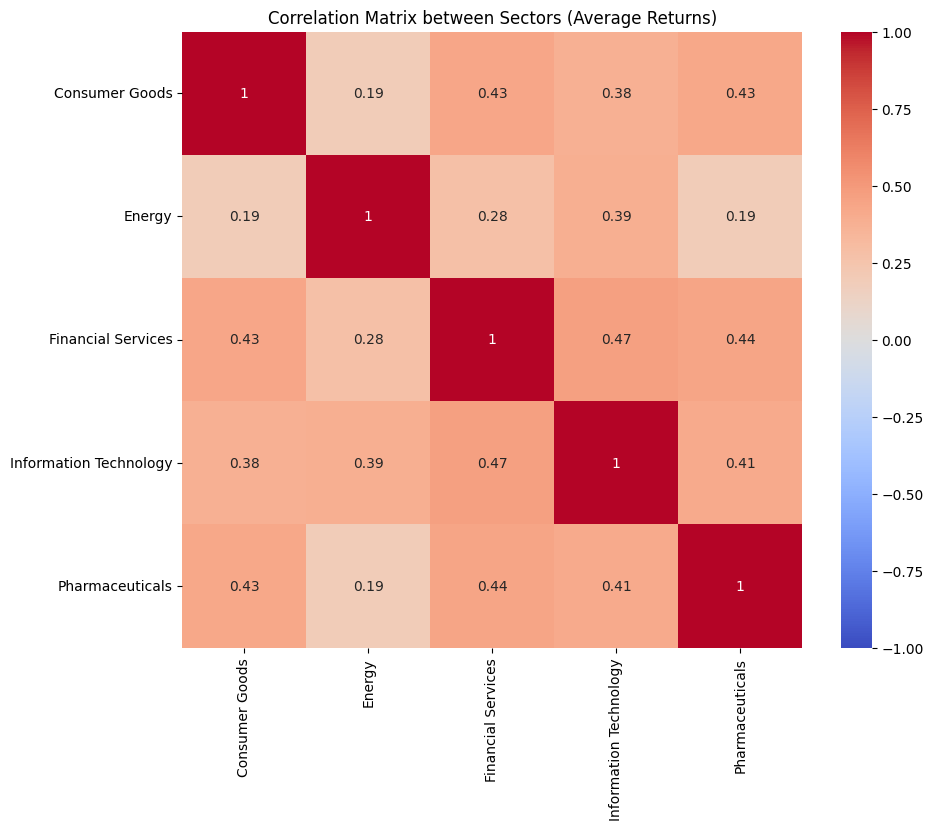}
    \caption{Correlation matrix between sectors based on average daily returns.}
    \label{fig:sector_correlation}
\end{figure}

To investigate relationships across different sectors, we compute a correlation matrix using average daily returns at the sector level. The resulting heatmap, shown in Figure~\ref{fig:sector_correlation}, reveals moderate positive correlations across most sectors. Notably, Financial Services and Information Technology show stronger co-movement with other sectors, while Energy exhibits relatively weaker correlations. These inter-sector dynamics highlight the presence of systemic links, reinforcing the importance of capturing sectoral structure in stress propagation models.

\subsection{Time Series Analysis}

To examine the time series properties of the return data, we apply the Augmented Dickey-Fuller (ADF) test to each stock's return series. The test results yield low p-values across all stocks, providing strong statistical evidence to reject the null hypothesis of a unit root. This indicates that the return series are stationary, making them suitable for use in modeling frameworks that assume stationarity like PCA. While AEs and VAEs do not inherently assume stationarity of the input data, it can improve the reliability of the mapping and leads to better performance.

To analyze volatility dynamics, we apply the GARCH(1,1) model across representative stocks from each sector. This allows us to capture time-varying conditional volatility, which is a well-documented characteristic of financial return series. The estimated parameters, $\omega$, $\alpha$, and $\beta$, demonstrate that recent volatility (captured by $\alpha$) and past volatility (captured by $\beta$) both play significant roles in shaping current variance. A high $\alpha$ indicates a strong reaction to recent market shocks, while a high $\beta$ reflects prolonged periods of volatility.

Stocks from the energy and financial services sectors exhibit relatively higher volatility persistence, as indicated by their higher $\beta$ values, while companies in consumer goods and pharmaceuticals show comparatively lower persistence. This sector-level variation in volatility dynamics reinforces the importance of adopting sector-aware modeling frameworks in stress testing applications (see Table~\ref{tab:garch_summary}).

\subsection{Summary and Modeling Implications}

The analyses in this section highlight key structural features of the return data, stationarity, sectoral correlations, and volatility clustering, that inform our modeling choices. These properties justify the use of dimensionality reduction methods to extract latent factors and motivate the application of PCA, AE, and VAE for scenario generation in the stress testing framework.

\section{Results and Empirical Analysis}

The PCA-based stress testing pipeline uses raw daily return data over a rolling window of 252 trading days. For each window, PCA is applied using \texttt{sklearn.decomposition.PCA} with the number of components fixed at $d = 5$. No standardization is applied prior to decomposition. Stress scenarios are generated by perturbing individual PCs (by $\pm 2\sigma$) or multiple principal components (e.g., $[+2.0, -1.5, +1.0, +0.5, -0.5]$) $\overline{\sigma}$, followed by reconstruction using the inverse transformation via the PCA loading matrix.

The AE pipeline is trained on a rolling window of standardized daily returns, using a window length of 504 trading days (approximately two years). Returns are standardized using \texttt{StandardScaler}. The AE architecture consists of a symmetric, fully connected feedforward neural network with the following structure:
\begin{center}
\texttt{Input (25) $\rightarrow$ Dense(16, tanh) $\rightarrow$ Dense(5, tanh) $\rightarrow$ Dense(16, tanh) $\rightarrow$ Output (25)}
\end{center}
All hidden layers use tanh activation and the output layer uses linear activation function. The model is trained using the mean squared error (MSE) loss with early stopping based on validation loss. Optimization is performed using the Adam optimizer with a batch size of 32. After training, stress scenarios are generated either by perturbing individual latent factors (by $\pm 2\sigma$) or multiple factors (e.g., $[+2.0, -1.5, +1.0, +0.5, -0.5]$) $\overline{\sigma}$, and the perturbed latent representations are decoded to generate stressed return vectors. The standardized outputs are then inverse-transformed back to their original scale before portfolio-level evaluation.

The VAE pipeline follows the same data preprocessing and network architecture as the AE, with the key distinction that it models the latent space as a distribution. The encoder outputs both a mean vector $\mu \in \mathbb{R}^5$ and a log-variance vector $\log \sigma^2 \in \mathbb{R}^5$, defining a Gaussian posterior over the latent space. During training, latent samples are drawn using the reparameterization approach, as follows:
\begin{equation}
z = \mu + \sigma \odot \epsilon, \quad \epsilon \sim \mathcal{N}(0, I)
\end{equation}
The training objective combines the MSE reconstruction loss with the Kullback-Leibler (KL) divergence between the learned posterior and a standard Gaussian prior. Monte Carlo sampling is used to generate $M = 1000$ latent samples from the learned posterior in each window, which are decoded to obtain a distribution of stressed return vectors. These outputs are used to visualize the spread and skewness of portfolio-level stress responses without relying on deterministic delta metrics.

\subsection{PCA-Based Stress Testing Results}

We evaluate the impact of PCA-based stress scenarios by analyzing changes in portfolio-level risk metrics under both single-factor and multi-factor perturbations. The stress scenarios are designed by perturbing the top principal components derived from a rolling-window PCA applied to the raw return data.

\subsubsection{Single-Component Perturbation Results}

\begin{figure}[h]
\centering
\subfloat[Delta VaR (95\%) – Single Factor]{
\includegraphics[width=0.45\textwidth]{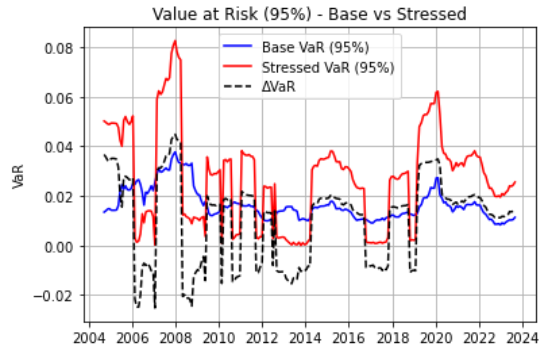}}
\hfill
\subfloat[Delta ES (95\%) – Single Factor]{
\includegraphics[width=0.45\textwidth]{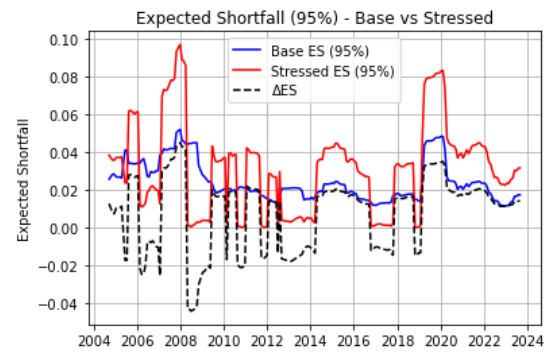}
}
\caption{Comparison of base and stressed portfolio risk under single-factor PCA stress.}
\label{fig:pca_single_stress}
\end{figure}

In the single-factor setup, we perturb the first principal component (PC1) by $\pm 2\sigma$, while keeping the remaining components fixed. This stress captures the effect of an isolated shift in the dominant linear risk factor. Figure~\ref{fig:pca_single_stress} presents the comparison of base and stressed portfolio risk across time for VaR and ES at the 95\% confidence level.

The results indicate a substantial increase in tail risk during major market downturns. Spikes in $\Delta$VaR and $\Delta$ES were especially prominent during the 2008 Global Financial Crisis, the 2018 correction, and the COVID-19 pandemic. The baseline metrics remain relatively stable across time, while the stressed risk metrics show significant deviations, emphasizing the sensitivity of the portfolio to dominant linear risk factors.

\subsubsection{Multi-Component Perturbation Results}

\begin{figure}[h]
\centering
\subfloat[Delta VaR (95\%) – Multi Factor]{
\includegraphics[width=0.45\textwidth]{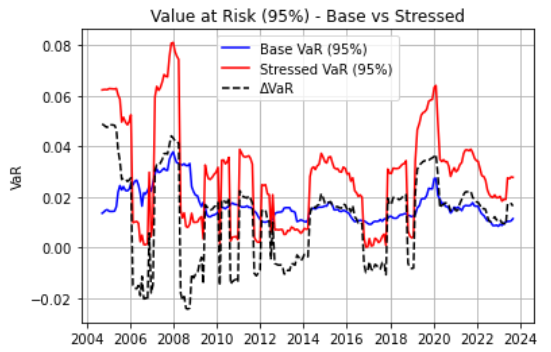}}
\hfill
\subfloat[Delta ES (95\%) – Multi Factor]{
\includegraphics[width=0.45\textwidth]{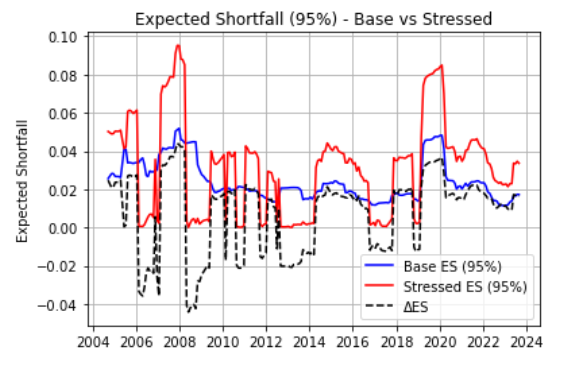}}
\caption{Impact of simultaneous multi-component perturbation on portfolio risk metrics.}
\label{fig:pca_multi_stress}
\end{figure}

We now simulate a systemic stress event by jointly perturbing the top five principal components using the vector $\Delta = [+2.0, -1.5, +1.0, +0.5, -0.5] \cdot \bar{\sigma}$, where $\bar{\sigma}$ is the standard deviation of each component over time. Figure~\ref{fig:pca_multi_stress} shows the resulting stressed risk metrics.

The multi-factor stress exhibits trends similar to the single-factor case, with elevated risk exposure observed during historical crisis periods. However, the magnitude and duration of the stress are slightly amplified, confirming that compounding shocks across components produce more persistent effects.

\subsubsection{Sector-Wise Impact Under Stress}

\begin{figure}[h]
\centering
\includegraphics[width=0.95\textwidth]{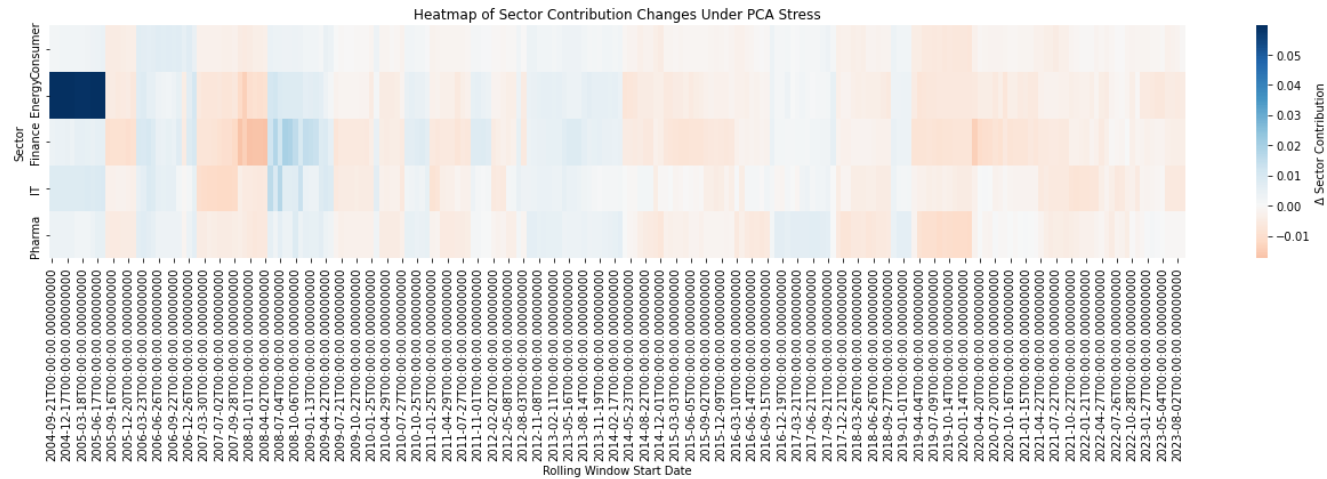}
\caption{Heatmap of sector-level contribution changes under PCA-based multi-factor stress.}
\label{fig:sector_heatmap}
\end{figure}

To assess how stress propagates across sectors, we examine the change in average sector-wise returns under the multi-component stress scenario. Figure~\ref{fig:sector_heatmap} displays the heatmap of sectoral return deltas over time.

The heatmap reveals distinct sectoral responses to stress scenarios. The financial sector often bears the highest negative impact, especially during crisis periods. Energy and Consumer Goods exhibit variable sensitivity, while Information Technology and Pharmaceuticals are relatively more stable. These observations validate the inclusion of sectoral structure in our stress testing framework.

\subsubsection{Component-Wise Alignment with Historical Events}

To validate the economic relevance of our stress scenarios, we examine the stress impact of each principal component during the 2008 Global Financial Crisis. Table~\ref{tab:pc_impact_gfc} summarizes the change in portfolio-level risk metrics due to perturbing each component individually.

\begin{table}[h]
\centering
\caption{Component-wise Stress Impact During the 2008 Global Financial Crisis}
\label{tab:pc_impact_gfc}
\begin{tabular}{lccc}
\toprule
\textbf{Principal Component} & $\Delta$VaR & $\Delta$ES & $\Delta$Drawdown \\
\midrule
PC1 & 0.0479 & 0.0479 & -0.6612 \\
PC2 & 0.0004 & 0.0004 & -0.0113 \\
PC3 & 0.0013 & 0.0013 & -0.0504 \\
PC4 & 0.0019 & 0.0019 & -0.0955 \\
PC5 & -0.0015 & -0.0015 & +0.0415 \\
\bottomrule
\end{tabular}
\end{table}

PC1 dominates the stress response, contributing the most to both tail risk and drawdown. Other components have relatively minor effects. This provides empirical evidence that PCA-based stress factors align well with real-world market shocks, reaffirming the validity of our approach.

\vspace{1em}
\noindent In summary, PCA-based stress testing provides a transparent and interpretable method to generate stress scenarios that align well with historical episodes of market distress. While effective in capturing linear co-movements, its limitations in modeling nonlinear dependencies motivate the need for more expressive frameworks, which are explored next using AE and VAE-based pipelines.

\subsection{AE-Based Stress Testing Results}

This section presents the results of applying AE-based stress testing to standardized daily returns across a 20-year time horizon. As discussed earlier, the AE pipeline captures nonlinear latent risk factors and simulates stress by perturbing these factors in the learned latent space. Both single-factor and multi-factor perturbations are used to assess portfolio vulnerabilities.

\subsubsection{Single-Latent Perturbation Results}

\begin{figure}[h]
\centering
\subfloat[Delta VaR (95\%) – Single Latent]{
\includegraphics[width=0.45\textwidth]{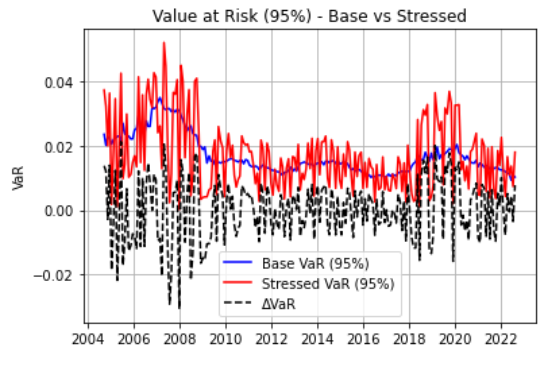}}
\hfill
\subfloat[Delta ES (95\%) – Single Latent]{
\includegraphics[width=0.45\textwidth]{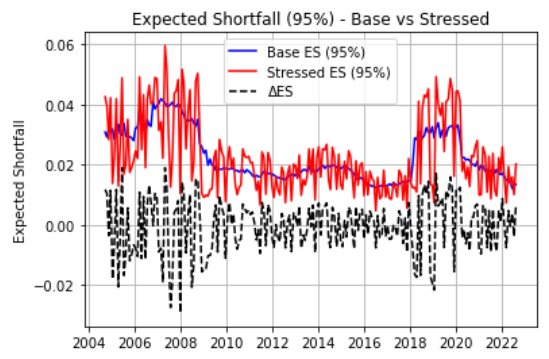}}
\caption{Portfolio risk changes under single-latent stress using AE.}
\label{fig:ae_single_stress}
\end{figure}

We first perturb individual dimensions of the latent representation obtained via the trained AE. Figure~\ref{fig:ae_single_stress} shows the change in portfolio risk metrics, namely, VaR and ES, when a single latent dimension is perturbed by $\pm 2\sigma$.

The single-latent stress reveals episodic spikes in risk, most notably during the 2008 Global Financial Crisis, the 2018 correction, and the COVID-19 crash. While similar in trend to PCA-based stress, AE-based perturbations capture more subtle nonlinear responses, offering better expressiveness for real-world volatility.

\subsubsection{Multi-Latent Perturbation Results}

\begin{figure}[h]
\centering
\subfloat[Delta VaR (95\%) – Multi-Latent]{
\includegraphics[width=0.45\textwidth]{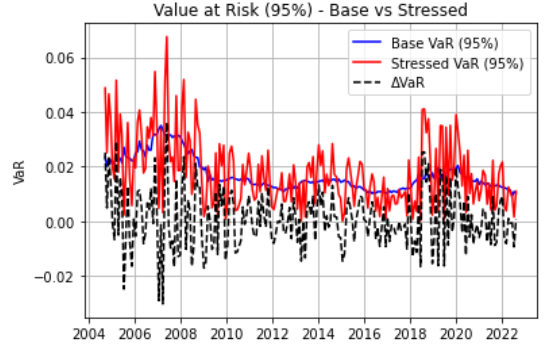}}
\hfill
\subfloat[Delta ES (95\%) – Multi-Latent]{
\includegraphics[width=0.45\textwidth]{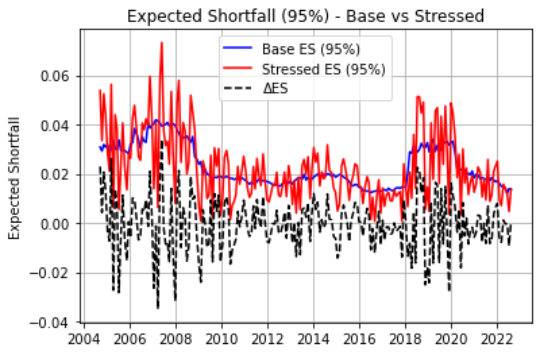}}
\caption{Impact of multi-latent perturbation on portfolio risk under AE-based stress.}
\label{fig:ae_multi_stress}
\end{figure}

We next apply a multi-latent stress by perturbing all latent dimensions simultaneously using the vector $\Delta = [+2.0, -1.0, +1.5, -0.5, +1.0] \cdot \bar{\sigma}$. This simulates a compound systemic shock encoded in the nonlinear latent space. The resulting $\Delta$VaR and $\Delta$ES are shown in Figure~\ref{fig:ae_multi_stress}.

The magnitude of stress response under multi-latent AE is consistently higher than that observed in the single-latent case. This highlights the nonlinearity of the learned representation, that is, simultaneous stress across dimensions leads to compounded portfolio effects that are not simply additive.

\subsubsection{Sector-Wise Sensitivity}

\begin{figure}[h]
\centering
\includegraphics[width=0.95\textwidth]{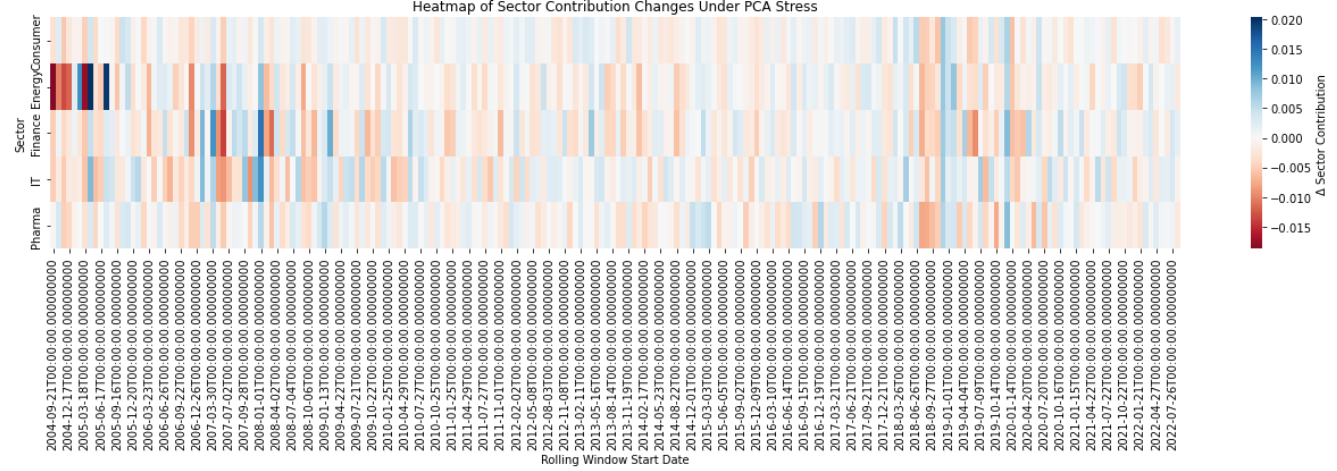}
\caption{Sector-wise contribution changes under AE-based multi-latent stress.}
\label{fig:ae_sector_heatmap}
\end{figure}

To interpret how different sectors respond to nonlinear stress factors, we compute the sector-level average return shifts under multi-latent stress. The heatmap in Figure~\ref{fig:ae_sector_heatmap} shows clear asymmetries across sectors and time.

The financial sector remains highly sensitive, with Energy and Consumer Goods also showing periodic vulnerability. The richer latent representation learned by the AE enables the model to identify complex patterns of sectoral stress propagation that may not be captured by PCA alone.

\subsubsection{Latent Factor Alignment with Historical Events}

To assess whether the latent dimensions learned by the AE capture meaningful sources of systemic risk, we evaluate their individual impact during the 2008 Global Financial Crisis. Each latent factor is perturbed in isolation and the resulting shifts in portfolio-level risk metrics, namely, $\Delta$VaR, $\Delta$ES and $\Delta$Drawdown are computed. Table~\ref{tab:ae_2008_impact} summarizes the change in portfolio-level risk metrics due to perturbing each component individually.

\begin{table}[h]
\centering
\caption{Component-wise Stress Impact During the 2008 Global Financial Crisis}
\label{tab:ae_2008_impact}
\begin{tabular}{lccc}
\toprule
\textbf{Latent Factor} & $\Delta$VaR & $\Delta$ES & $\Delta$Drawdown \\
\midrule
Z1 & $-0.0066$ & $-0.0061$ & $+0.2367$ \\
Z2 & $+0.0012$ & $+0.0011$ & $-0.3211$ \\
Z3 & $+0.0014$ & $+0.0012$ & $-0.3899$ \\
Z4 & $-0.0063$ & $-0.0062$ & $+0.2251$ \\
Z5 & $+0.0017$ & $+0.0016$ & $-0.4353$ \\
\bottomrule
\end{tabular}
\end{table}

The results indicate that certain latent directions, particularly $Z1$ and $Z4$, contribute significantly to risk amplification during crises, as evidenced by their large drawdown and expected shortfall effects. Others like $Z2$ and $Z3$ appear to capture milder variations, possibly reflecting sectoral or idiosyncratic patterns. 

This sensitivity analysis demonstrates that the latent space learned by the AE is not merely a statistical compression, but encodes directions that meaningfully reflect market stress. Such interpretability strengthens the case for using AE-based methods in real-world stress testing frameworks.

\subsection{VAE-Based Monte Carlo Stress Testing Results}

The VAE framework enables a fundamentally different approach to stress testing by learning a probabilistic latent representation of return dynamics. Unlike PCA and AE, where scenarios are generated through direct perturbations in latent space, the VAE allows for sampling from a learned distribution, offering a Monte Carlo-based mechanism to explore a broader range of plausible stress scenarios.

While the VAE architecture can also be used in a deterministic mode, mimicking AE-style latent perturbations, has the key advantage lying in its ability to simulate return trajectories under uncertainty. By drawing multiple samples from the latent posterior distribution and decoding them into stressed return vectors, we obtain a distribution of portfolio outcomes rather than a single trajectory.

\begin{figure}[h]
\centering
\includegraphics[width=0.6\textwidth]{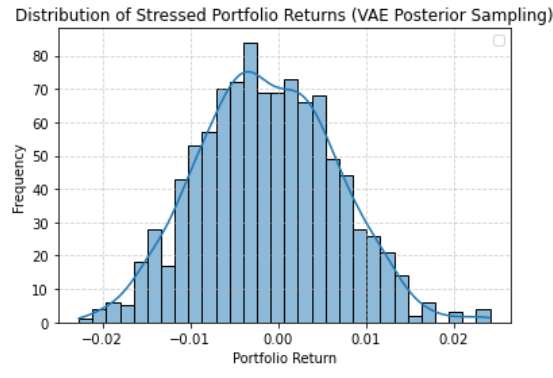}
\caption{Distribution of portfolio returns under VAE posterior sampling.}
\label{fig:vae_mc_returns}
\end{figure}

Figure~\ref{fig:vae_mc_returns} shows the resulting distribution of stressed portfolio returns generated from posterior latent samples. The bell-shaped curve closely resembles a Gaussian profile, but with slight asymmetry and heavier tails, suggesting the VAE has captured important aspects of return uncertainty beyond that of purely linear or deterministic models.

Although conventional risk metrics such as $\Delta$VaR or $\Delta$ES are not directly applicable under this sampling-based setup, the framework provides a powerful mechanism for distributional stress testing, enabling institutions to analyze the entire risk surface rather than a few point estimates.

\subsection{Cross-Method Comparison and Discussion}

The three stress testing pipelines, PCA, AE, and VAE, offer distinct advantages in modeling latent risk factors and evaluating portfolio resilience.

PCA provides a transparent and efficient baseline. Its linear components are easy to interpret and align well with broad market or sector-level movements, making attribution straightforward. However, its inability to model nonlinear dependencies can limit its expressiveness, especially in complex financial environments.

AE addresses this limitation by learning nonlinear latent factors that capture deeper interactions across assets. The stress scenarios generated reflect more flexible patterns of co-movement, particularly evident during crisis periods. While more powerful, AE sacrifices some interpretability due to the black-box nature of neural encoders.

VAE extends this further by introducing uncertainty into the latent space. Stress scenarios are drawn from the learned posterior, enabling Monte Carlo-style simulation of systemic shocks. This probabilistic approach is useful for exploring risk distributions under uncertainty, though it offers less direct attribution and requires greater computational effort.

Across all methods, stress sensitivity during historical crises, such as 2008 and 2020, was consistently observed. PCA highlighted these through sharp $\Delta$VaR and $\Delta$ES jumps, while AE and VAE captured more distributed and nuanced latent shifts. Sectoral stress impacts also varied: AE revealed more balanced effects, whereas PCA concentrated stress in dominant sectors.

In summary, PCA offers interpretability, AE captures richer structure, and VAE enables probabilistic stress simulation. Together, they provide a complementary suite of tools for robust, data-driven financial stress testing.

\section{Conclusion and Future Work}

This work presents a machine learning-based framework for stress testing sectoral portfolios in the Indian financial market. Our approach integrates linear and nonlinear dimensionality reduction techniques, namely, PCA, AE and VAE, within a modular pipeline for scenario generation and portfolio risk evaluation. By perturbing or sampling from learned latent representations, we construct stress scenarios that simulate both targeted shocks and systemic uncertainty.

The PCA pipeline offers a transparent baseline, allowing interpretable analysis of stress effects aligned with known market events. The AE framework captures richer nonlinear dependencies across sectors, revealing latent risk factors that are not easily isolated in linear models. The VAE model builds on this by introducing distributional uncertainty through Monte Carlo sampling, offering a probabilistic perspective on portfolio stress.

Our results show consistent sector-level vulnerabilities across all three pipelines, particularly in Financial Services and Energy during periods of elevated market stress. The framework not only reproduces historical crises but also demonstrates its adaptability across different modeling paradigms, laying the groundwork for more flexible, machine learning-driven stress testing methodologies.

This framework serves as a foundation for further research in applying machine learning to financial risk management. Future work can explore alternative architectures for scenario generation and latent factor modeling, including attention-based networks, sequence models, and deep generative methods such as GANs or normalizing flows. These techniques offer the potential to capture more complex temporal and cross-sectional structures in market behavior.

The modular structure of our pipeline also allows for extensions to multi-asset portfolios, the inclusion of macroeconomic and sentiment data, and experimentation with adaptive models under regime-switching dynamics. By combining the flexibility of modern ML architectures with domain-specific insights from finance, future studies can build more robust, forward-looking systems for stress testing and systemic risk evaluation.

\section*{Data Availability Statement}

The data used for the study is publicly available and will be made available upon request.

\section*{Conflict of Interest Statement}

The authors declare no conflict of interest.

\bibliographystyle{elsarticle-num}

\bibliography{Biblio}

\appendix
\section*{Appendix A: Sector-wise Stock List}

\begin{table}[H]
\centering
\caption{List of companies selected by sector and corresponding ticker symbols}
\label{tab:sector_companies}
\renewcommand{\arraystretch}{0.8}  
\begin{tabular}{|c|c|c|}
\hline
\textbf{Sector} & \textbf{Company} & \textbf{Ticker Symbol} \\
\hline
\multirow{5}{*}{Financial Services} 
& HDFC Bank & HDFCBANK \\
& ICICI Bank & ICICIBANK \\
& State Bank of India & SBIN \\
& Axis Bank & AXISBANK \\
& Kotak Mahindra Bank & KOTAKBANK \\
\hline
\multirow{5}{*}{Information Technology} 
& Tata Consultancy Services & TCS \\
& Infosys & INFY \\
& Wipro & WIPRO \\
& HCL Technologies & HCLTECH \\
& Tech Mahindra & TECHM \\
\hline
\multirow{5}{*}{Consumer Goods} 
& Hindustan Unilever & HINDUNILVR \\
& Nestlé India & NESTLEIND \\
& ITC Limited & ITC \\
& Dabur India & DABUR \\
& Britannia Industries & BRITANNIA \\
\hline
\multirow{5}{*}{Energy} 
& Reliance Industries & RELIANCE \\
& NTPC Limited & NTPC \\
& Indian Oil Corporation & IOC \\
& Bharat Petroleum Corporation & BPCL \\
& Tata Power & TATAPOWER \\
\hline
\multirow{5}{*}{Pharmaceuticals} 
& Sun Pharmaceutical Industries & SUNPHARMA \\
& Dr. Reddy’s Laboratories & DRREDDY \\
& Cipla & CIPLA \\
& Lupin Limited & LUPIN \\
& Aurobindo Pharma & AUROPHARMA \\
\hline
\end{tabular}
\end{table}

\section*{Appendix B: GARCH(1,1) Model Coefficients for Selected Companies}
\begin{table}[H]
\centering
\caption{GARCH(1,1) Model Coefficients for Selected Companies}
\label{tab:garch_summary}
\begin{tabular}{|l|c|c|c|}
\hline
\textbf{Company} & $\omega$ & $\alpha$ & $\beta$ \\
\hline
Britannia & 0.2291 & 0.0824 & 0.8340 \\
Dabur & 0.1756 & 0.1535 & 0.8149 \\
BPCL & 55.9961 & 0.0082 & 0.0000 \\
Axis Bank & 0.0964 & 0.0768 & 0.9077 \\
Infosys & 0.5283 & 0.1633 & 0.6948 \\
Cipla & 0.3283 & 0.0958 & 0.8022 \\
\hline
\end{tabular}
\end{table}

\end{document}